\documentclass[a4paper,10pt]{article}

\usepackage[utf8]{inputenc}
\usepackage[T1]{fontenc}
\usepackage{graphicx}               % 図の読み込み
\usepackage{setspace}               % 行間調節用パッケージ
\usepackage{cite}                   % 引用スタイル
\usepackage{amsmath, amssymb,amsfonts, amsthm}  % 数学パッケージ
\usepackage{physics}                % 物理パッケージ(https://qiita.com/HelloRusk/items/ce9f49e9b3fc0344ae23)
\usepackage{lineno}
\usepackage{color}                  % 色付け用パッケージ
\usepackage{authblk}                % 著者の所属を記載するパッケージ

\newcommand*\patchAmsMathEnvironmentForLineno[1]{
  \expandafter\let\csname old#1\expandafter\endcsname\csname #1\endcsname
  \expandafter\let\csname oldend#1\expandafter\endcsname\csname end#1\endcsname
  \renewenvironment{#1}
  {\linenomath\csname old#1\endcsname}
{\csname oldend#1\endcsname\endlinenomath}}
\newcommand*\patchBothAmsMathEnvironmentsForLineno[1]{
  \patchAmsMathEnvironmentForLineno{#1}
\patchAmsMathEnvironmentForLineno{#1*}}
\AtBeginDocument{
  \patchBothAmsMathEnvironmentsForLineno{equation}
  \patchBothAmsMathEnvironmentsForLineno{align}
  \patchBothAmsMathEnvironmentsForLineno{flalign}
  \patchBothAmsMathEnvironmentsForLineno{alignat}
  \patchBothAmsMathEnvironmentsForLineno{gather}
  \patchBothAmsMathEnvironmentsForLineno{multline}
}
\title{Phase Relationship between Spinal Motion and Limb Support Determines High-speed Running Performance in a Cheetah Model with Asymmetric Spinal Stiffness}

\author[1]{Tomoya~Kamimura}
\author[2]{Yuya~Oshita}
\author[1]{Mau~Adachi}
\author[3]{Yuichi~Ambe}
\author[2]{Akihito~Sano}
\author[4]{Naomi~Wada}
\author[5]{Fumitoshi~Matsuno}
\author[1]{Shinya~Aoi}

\affil[1]{\small Department of Mechanical Science and Bioengineering, Graduate School of Engineering and Science, The University of Osaka, Osaka, Japan}
\affil[2]{\small Department of Electrical and Mechanical Engineering, Nagoya Institute of Technology, Aichi, Japan}
\affil[3]{\small Department of Electrical, Systems, and Control Engineering, Graduate School of Advanced Science and Engineering, Hiroshima University, Hiroshima, Japan}
\affil[4]{\small Laboratory of System Physiology, Joint Faculty of Veterinary Medicine, Yamaguchi University, Yamaguchi, Japan}
\affil[5]{\small Department of Electronics and Information Systems Engineering, Osaka Institute of Technology, Osaka, Japan}

\date{}

\begin{document}

\maketitle
\setstretch{1.4} % ページ全体の行間を設定
% \linenumbers % 行番号を表示

\begin{abstract}
%   チーターは走行中に背骨を大きく曲げ伸ばしすることが特徴である．
%   本研究では，チーターからインスパイアされた非対称な背骨剛性を持つシンプルな動力学モデルを用いて，背骨の曲げ伸ばし運動と前後肢の接地期の位相関係が走行に与える影響を調べた．
%   様々な背骨剛性に対してモデルの周期解を数値的に探索した結果，チーターと同様に後肢接地期のあとに背骨が伸ばされる解と，逆に後肢接地期のあとに背骨が曲げられる解が得られた．
%   背骨剛性が非対称な場合は特に，背骨の曲げ伸ばしに対して前後肢の接地期がチーターと同様の位相関係で発生することで，移動速度を維持しながら床反力を抑制できることが明らかになった．
%   本研究は，チーターの高速走行に対して力学的な理解を与え，生物学の理解を深めるだけでなく，ロボットなどの工学的な応用の可能性を示すものである．

Cheetahs are characterized by large spinal flexion and extension during high-speed running, yet the dynamical role of the phase relationship between spinal motion and limb support remains unclear. We aimed to clarify how this phase relationship affects running performance, focusing on the effect of asymmetric spinal stiffness. Using a simple planar cheetah model with asymmetric torsional spinal stiffness, we numerically searched for periodic bounding solutions over a range of stiffness parameters and compared their ground reaction forces, horizontal velocities, and stability. We obtained both cheetah-like solutions, in which the spine extends after hindlimb liftoff and flexes after forelimb liftoff, and non-cheetah-like solutions, in which the spine flexes after hindlimb liftoff and extends after forelimb liftoff. Under asymmetric spinal stiffness, cheetah-like solutions reduced ground reaction forces while maintaining horizontal velocity more effectively than non-cheetah-like solutions. The phase relationship between spinal motion and stance timing is a key determinant of high-speed running performance. These findings provide a dynamical understanding of cheetah locomotion and suggest design principles for spined legged robots. 
\end{abstract}


%-----------------------------------------------------------------------

\section{Introduction}
% 地上で最も高速に走行するチーターは，大きく体幹を曲げ伸ばしするのが特徴であり，その効果は古くから運動学・動力学の視点から様々指摘されている\cite{Hildebrand1959-gw,Bertram2009-yp,Shield2023-ch}．
% 例えば，脚を前後に大きく曲げ伸ばしすることで，ストライド長を伸ばすことができる．
% 加えて，体幹部の筋肉（特に背側の筋肉）を追加のアクチュエータとして活用し，多くのエネルギーを走行中に注入できる．
% これまでの指摘は主に観測に基づくものであったが，動物の運動は身体・神経・環境の相互作用からなる複雑な力学現象であり，その原理をすべて観察から理解することは困難である．
% そこで，シンプルな力学モデルを用いて運動を構成論的に理解しようとするアプローチが用いられてきた\cite{Alexander1988-uj,Swanstrom2005-cq,Bertram2009-yp,Aoi2017-kj,Ambe2018-dt,Toeda2019-le,Polet2019-iq,Polet2020-uf,Polet2022-xw,Adachi2025-zg,Yaegashi2025-ko}．
% 例えば走行はバネと質点からなるSpring-Loaded Inverted Pendulum (SLIP)モデルによってその運動がよく近似でき\cite{Schmitt2000-ka,Schmitt2000-kl,Full1999-xd,Blickhan1993-ei,Farley1993-fw}，近年では，SLIPモデルを拡張して，4足動物の運動の解析も行われている\cite{Poulakakis2006-ww,Tanase2015-bt,Gan2015-et,Gan2018-bj,Gan2018-kk,Yamada2022-mn}．
% 動物のロコモーションは脚だけでなく全身で実現される．
% チーターの体幹部の曲げ伸ばしを再現するモデルも構築され，解析が行われてきた\cite{Cao2013-gh,Wang2017-dg,Kamimura2015-ku,Kamimura2018-dx,Yesilevskiy2018-sw,Adachi2020-oj}．

Cheetahs, the fastest terrestrial runners, are characterized by large flexion--extension of the spine, and its effects have long been discussed from kinematic and dynamic perspectives~\cite{Hildebrand1959-gw,Bertram2009-yp,Shield2023-ch}.
For example, large flexion and extension of the body can increase stride length.
In addition, spinal muscles, especially epaxial muscles, can act as additional actuators and inject substantial energy during running.
Although these effects have mainly been inferred from observation, animal locomotion is a complex dynamical phenomenon arising from interactions among the body, nervous system, and environment, and it is difficult to understand all of its principles from observation alone.
Accordingly, constructive approaches based on simple models have been widely used to investigate locomotion~\cite{Alexander1988-uj,Swanstrom2005-cq,Bertram2009-yp,Aoi2017-kj,Ambe2018-dt,Toeda2019-le,Polet2019-iq,Polet2020-uf,Polet2022-xw,Adachi2025-zg,Yaegashi2025-ko}.
For instance, running can be well approximated by the Spring-Loaded Inverted Pendulum (SLIP) model, which consists of a spring and a mass point~\cite{Schmitt2000-ka,Schmitt2000-kl,Full1999-xd,Blickhan1993-ei,Farley1993-fw}, and this model has recently been extended to analyze quadrupedal locomotion~\cite{Poulakakis2006-ww,Tanase2015-bt,Gan2015-et,Gan2018-bj,Gan2018-kk,Yamada2022-mn}.
Because animal locomotion is achieved by the whole body rather than by legs alone, models reproducing cheetah spinal flexion--extension have also been developed and analyzed~\cite{Cao2013-gh,Wang2017-dg,Kamimura2015-ku,Kamimura2018-dx,Yesilevskiy2018-sw,Adachi2020-oj}.
In our previous study, we showed that appropriate spinal motion passively suppresses excessive ground reaction force (GRF), thereby contributing to improved horizontal velocity and stability~\cite{Kamimura2022-fa}.

% 走行中，接地期において並進速度が加減する．具体的には，接地期前半には減速し，後半に加速する．
% ウマは前肢接地期の後にのみ足が全て地面から離れる飛翔期を持つ．後肢接地期後半に前肢が接地するので，後肢による加速が阻害される．
% これに対して，チーターはそのような阻害を受けない．具体的には，脊柱が伸ばされ前後肢が離れるextended flight，forelimb stance，脊柱が曲げられて前後肢が近づくgathered flight，hindlimbの4つのフェーズを繰り返し(Fig.~\ref{fig:cheetah_gait})，前肢接地期後と後肢接地期後の両方に飛翔期を持つため，後肢接地期の加速を十分に活用し高速走行を実現している\cite{Bertram2009-yp}．
% 加えて，後肢接地期後半には体幹部の筋肉も用いて加速を行うことによって，チーターは他に類を見ない速度を実現できる．
% これまで著者らが用いていたチーターのモデル\cite{Kamimura2022-fa}では，チーターと同様に前肢接地後と後肢接地後にそれぞれ飛翔期を持つ周期解が得られていた．
% ただし，その中にはチーターと同様に後肢離地後に体幹を伸ばして前肢離地後に体幹を曲げる解に加えて，チーターと違って後肢離地後に体幹を曲げて前肢離地後に体幹を伸ばす解，すなわち，体幹の曲げ伸ばしに対する前肢後肢の離地順が逆になる解，が含まれていた．
% これらの解の間には安定性に違いが見られたものの，最大床反力や進行速度など他のパフォーマンスに大きな差が見られず，チーターが前者の歩容を用いる理由を十分に説明できていなかった．

\begin{figure}
  \centering
  \includegraphics[scale=1.0]{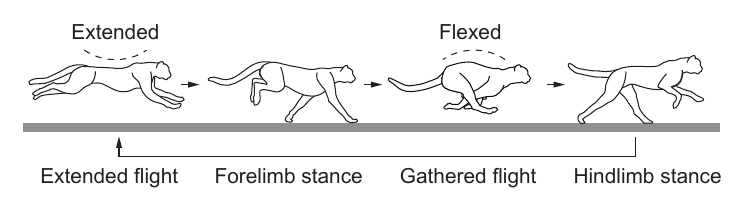}
  \caption{%
    % チーターの歩容のシークエンス．背骨が伸ばされ前後肢が離れるextended flight，forelimb stance phase，背骨が曲げられて前後肢が近づくgathered flight，hindlimb stanceの順となる．
    Sequence of the cheetah gait: extended flight (the spine is extended and the fore and hind limbs are apart), forelimb stance, gathered flight (the spine is flexed and the fore and hind limbs are close), and hindlimb stance.
  }
  \label{fig:cheetah_gait}
\end{figure}

During running, horizontal velocity decreases in the first half of stance and increases in the second half.
Horses have a flight phase, in which all feet leave the ground, only after the forelimb stance.
Because the forelimbs contact the ground in the latter half of hindlimb stance, hindlimb-driven acceleration is inhibited.
In contrast, cheetahs are not subject to this inhibition.
Specifically, their gait consists of four phases---extended flight (with the spine extended and the fore and hind limbs separated), forelimb stance, gathered flight (with the spine flexed and the fore and hind limbs close together), and hindlimb stance (Fig.~\ref{fig:cheetah_gait})---and includes flight phases after both forelimb and hindlimb stance~\cite{Bertram2009-yp}.
Therefore, cheetahs can fully exploit hindlimb-driven acceleration and achieve extraordinarily high-speed running.
Moreover, by using spinal muscles to accelerate in the latter half of hindlimb stance, cheetahs can achieve unmatched speed.
In the cheetah model used in our previous study~\cite{Kamimura2022-fa}, we obtained periodic solutions that, as in cheetahs, had flight phases after both forelimb and hindlimb stance.
However, these included two types of solutions: cheetah-like solutions in which the spine is extended after hindlimb liftoff and flexed after forelimb liftoff, and cheetah-unlike solutions in which the spine is flexed after hindlimb liftoff and extended after forelimb liftoff.
Although stability differed between these solutions, no other large differences were found in performance criteria such as maximum GRF and horizontal velocity; thus, this model could not fully explain why cheetahs use the former gait.

% そこで本研究では，脊柱剛性の非対称性と体幹の曲げ伸ばしに対する前後肢の接地タイミングに着目して，この問題を解決する．
% まず，チーターなどの4足動物では，脊柱剛性は曲げと伸ばしで非対称であると考えられる．
% これは，重力に抗して立ち内臓を支持する必要があることに加え，後肢離地時に脊柱の伸筋である軸上筋が加速に活用されることにも起因する．
% % そして，後肢加速を十分に活用するには，前肢接地後と後肢接地後の両方に飛翔期を持つことが重要であると指摘されている\cite{Bertram2009-yp．
% さらに先行研究\cite{Kamimura2022-fa}では，飛翔期そのものよりも接地期における脊柱の運動が走行に重要な役割を果たしており，特に，接地期における脊柱の適切な姿勢が床反力の低減につながることを示した．
% ただし，モデルの脊柱剛性を対称としたため，脊柱の曲げと伸ばしの力学的な差がほとんどなく，脊柱運動に対する前後肢の接地離地のタイミング，すなわち前後肢の離地後に脊柱を曲げるか伸ばすかの違いが走行に与える影響は十分に明らかになっていなかった．
% そこで本研究では，非対称な脊柱剛性を用いて，脊柱運動と前後肢の接地期の位相関係が走行に与える影響を調べる．

In this study, we address this issue by focusing on asymmetric spinal stiffness and on the timing of forelimb and hindlimb stance relative to spinal flexion--extension.
First, in quadrupeds such as cheetahs, spinal stiffness is considered asymmetric between flexion and extension.
This asymmetry may arise not only because animals must stand against gravity and support their internal organs, but also because they accelerate at hindlimb liftoff using epaxial muscles, which are spinal extensors.
It has also been pointed out that, to fully utilize hindlimb-driven acceleration, it is important to have flight phases after both forelimb and hindlimb stance~\cite{Bertram2009-yp}.
Furthermore, our previous model~\cite{Kamimura2022-fa} showed that spinal motion during stance, rather than flight itself, plays an important role in running and that an appropriate spinal posture during stance reduces GRF.
However, because spinal stiffness was assumed to be symmetric in the previous model, the mechanical difference between spinal flexion and extension was minimal, and the effects on running of differences in forelimb/hindlimb stance--liftoff timing relative to spinal motion (i.e., whether the spine is flexed or extended after limb liftoff) remained unclear.
Therefore, in this study, we investigate how the phase relationship between spinal motion and forelimb/hindlimb stance affects running by using asymmetric spinal stiffness.

\section{METHOD}

\subsection{Model}
% 背骨の柔軟性を持つモデルのバウンド歩容を調べるため，先行研究\cite{Kamimura2022-fa}と同様に2つの剛体(bodies 1 and 2)からなる2次元のモデルを用いた(Fig.~\ref{fig:model})．
% 左右脚は同時に動くと仮定して，モデルは2つの質量の無視できるバネからなる前後脚(legs 1 and 2)を持つ．
% 体幹の柔軟性を表現するために2つの剛体がねじりバネ$k_\mathrm{t}$を含む体幹部のジョイントで接続されている．
% 全体の質量中心の位置を$(x,y)$，前半身のピッチ角を$\theta_1$，後半身のピッチ角を$\theta_2$とする．
% ただし，これらの角度は反時計回りを正とする．
% ここで，体幹部の平均ピッチ角$\theta=\frac{\theta_1+\theta_2}{2}$と前後胴体の相対ピッチ角$2\phi=\theta_1-\theta_2$を定義する．
% 体幹を反らす（背骨が`$\vee$'型になる）姿勢と曲げる姿勢（背骨が`$\wedge$'型になる）がそれぞれ$\phi>0$，$\phi<0$となる．
% 体幹剛性の非対称性を再現するため，$\phi=0$をねじりばねの平衡点とし，$k_\mathrm{t}$を平衡点を境にしてステップ関数で切り替え，定数$k_0$と乗数$\kappa$を用いて以下のように定義する．
% \begin{align*}
%   k_\mathrm{t}=
%   \begin{cases}
%     \kappa k_0 & (\phi > 0)\\
%     k_0 & (\phi \leq 0)
%   \end{cases}
% \end{align*}
% ここで，モデルはチーターと同様に，体幹を反らす側に硬いと仮定する．
% すなわち$\kappa \geq 1$とする．
% モデルの前後のパーツは対称であると仮定する．それぞれの剛体の質量と質量中心周りの慣性モーメントはそれぞれ$m$と$J$である．それぞれの剛体の長さは$2r$である．
% 質量中心と足の付根の距離は$d$である．
% 重力加速度は$g$である．脚バネの自然長とバネ定数はそれぞれ$l_0$と$k$である．
% 鉛直下向きから反時計回りを正として脚バネ$i$ ($i=1,2$)のなす角度を$\gamma_i$とし，長さを$l_i$とする．
% 脚バネ$i$が空中にあるとき，その長さは$l_0$を，姿勢は接地角$\gamma_i^\mathrm{td}$を保つ．
% 脚$i$の接地期において，時計回りを正として，body $i$と脚$i$のなす相対角度を$\psi_i = \frac{\pi}{2} + \theta_i - \gamma_i$とする．

\begin{figure}
  \centering
  \includegraphics[scale=1.0]{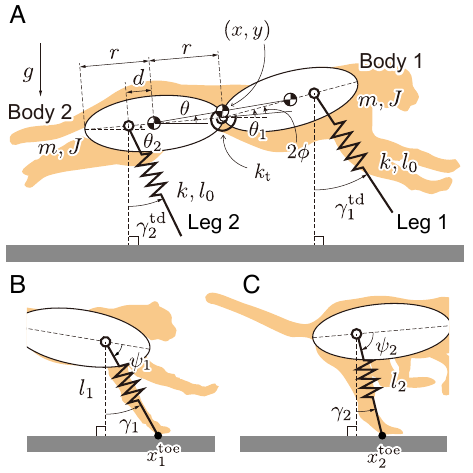}
  \caption{Simple model of cheetah.
  % Torsional spring in body joint changes its stiffness $k_\mathrm{t}$ according to body flexion angle $\phi$.
  (A) Flight phase. (B) Forelimb (Leg~1) stance. (C) Hindlimb (Leg 2) stance.}
  \label{fig:model}
\end{figure}

We used a simple cheetah model shown in Fig.~\ref{fig:model}.
This model is constructed based on the model used in~\cite{Kamimura2022-fa}, in which two rigid bodies (Bodies 1 and 2) are connected by a joint including torsion spring to represent the flexibility of the spine.
We assumed that the left and right limbs move simultaneously and represented the forelimb and hindlimb as massless prismatic springs (Legs 1 and 2).
The model is constrained in the sagittal plane.
% The equilibrium point of the body spring is the posture when the anterior and posterior torso are parallel.
% The fore and hindlimbs are represented by massless linear prismatic springs.
% The left and right legs are assumed to move simultaneously, and the model is constrained in the sagittal plane.
The position of the center of mass (COM) is given as $(x,y)$, the pitch angles of the fore body and hind body as $\theta_1$ and $\theta_2$, respectively, where the counterclockwise rotation is defined as positive.
The average pitch angle of the rigid bodies is defined as $\theta=\frac{\theta_1+\theta_2}{2}$, and the relative pitch angle of the fore and hind bodies is defined as $2\phi=\theta_1-\theta_2$.
We define the extended posture of the spine (spine in a `$\vee$' shape) and the flexed posture (spine in a `$\wedge$' shape) as $\phi>0$ and $\phi<0$, respectively.
To reproduce the asymmetry of spinal stiffness, we set $\phi=0$ as the equilibrium point of the torsional spring, and defined spring constant $k_\mathrm{t}$ using a step function with a constant $k_0$ and a multiplier $\kappa$ as follows:
\begin{align}
  k_\mathrm{t}=
  \begin{cases}
    \kappa k_0 & (\phi > 0) \\
    k_0 & (\phi \leq 0)
  \end{cases}
\end{align}
As in cheetahs, we assumed that the spine is stiffer in extended posture, i.e., $\kappa \geq 1$.
The fore and hind body segments are assumed to be symmetric.
The mass and moment of inertia about the COM of each rigid body are $m$ and $J$, respectively, and the length of each rigid body is $2r$.
The distance between the COM and the leg root is $d$.
The gravitational acceleration is $g$.
The nominal length and spring constant of the legs are $l_0$ and $k$, respectively.
The length of Leg $i$ ($i=1,2$) is $l_i$.
The angle of Leg $i$ from the vertical downward direction is $\gamma_i$, where counterclockwise rotation is taken as positive.
When Leg $i$ is in the air, its length is fixed at $l_0$, and its posture is fixed at the touchdown angle $\gamma_i^\mathrm{td}$.
During a stance of Leg $i$, the relative angle between Body $i$ and Leg $i$ is defined as $\psi_i( = \frac{\pi}{2} + \theta_i - \gamma_i)$, where clockwise rotation is taken as positive.

% 足先が地面に接地する瞬間，足先は地表面に拘束され，摩擦のないピンジョイントとして振る舞う．脚が縮んで再び自然長になる瞬間に離地し，脚の角度は瞬時に接地角$\gamma_i^\mathrm{td}$に戻る．
% このモデルは摩擦やダンパーの散逸項を持たず，バネ定数の切り替わりはバネが自然長になったとき（$\phi=0$）に行われ，接地・離地は脚バネが自然長のときに発生するため，エネルギー保存系である．

When the tip of the leg reaches the ground, it is constrained on the ground and behaves as a frictionless pin joint.
When the stance leg returns to its nominal length after compression, the leg angle immediately returns to the touchdown angle $\gamma_i^\mathrm{td}$.
This model conserves energy because the body spring-constant switching occurs when the spring reaches its equilibrium point ($\phi=0$), touchdown and liftoff occur at the nominal length, and the model has no dissipative structure, such as friction or damping.

% 運動方程式は，ラグランジュの運動方程式から以下のように導かれる~\cite{Kamimura2022-fa}．

% 〜

% なお，脚バネ$i$は遊脚期に$l_i=l_0$となるため，接地している間しかバネ力は発生しない．

% チーターの計測データ~\cite{Kamimura2021-ep}から，物理パラメータを以下のように決定した．$m = 19$~kg, $J = 0.53$~kgm$^2$, $r = 0.29$~m, $l_0 = 0.69$~m, and $d = 0.06$~m．体幹部の曲げ伸ばしと歩容周期の観点から，チーターと類似した走行動作を再現するために，$k_0 = 100$~Nm/rad と $k = 15000$~N/m を用いた．

The equations of motion of this model are given by~\cite{Kamimura2022-fa}.
\begin{align}
  M(q)\ddot{q} + h(q, \dot{q}) + G(q) = 0,
\end{align}
where $q=[x \ y \ \theta \ \phi]^\top$,
\begin{align*}
  M(q)&=
  \begin{bmatrix}
    2m & 0 & 0 & 0\\
    0 & 2m & 0 & 0\\
    0 & 0 & 2J+2mr^2\cos^2\phi & 0\\
    0 & 0 & 0 & 2J+2mr^2\sin^2\phi
  \end{bmatrix},\\
  h(q,\dot{q})&=
  \begin{bmatrix}
    0\\
    0\\
    4mr^2\dot{\theta}\dot{\phi}\cos\phi\sin\phi\\
    -2mr^2(\dot{\theta}^2+\dot{\phi}^2)\cos\phi\sin\phi
  \end{bmatrix},\\
  G(q)&=
  \begin{bmatrix}
    0\\
    2mg\\
    0\\
    4 k_\mathrm{t}\phi
  \end{bmatrix} + k(l_1-l_0)
  \begin{bmatrix}
    \partial l_1/\partial x\\
    \partial l_1/\partial y\\
    \partial l_1/\partial \theta\\
    \partial l_1/\partial \phi
  \end{bmatrix} + k(l_2-l_0)
  \begin{bmatrix}
    \partial l_2/\partial x\\
    \partial l_2/\partial y\\
    \partial l_2/\partial \theta\\
    \partial l_2/\partial \phi
  \end{bmatrix},
\end{align*}
\begin{align*}
  l_i =
  \begin{cases}
    l_0, & \text{when Leg $i$ is in the air}\\
    \begin{aligned}
      &\left[\left\{x+(-1)^i(r+d)\cos\theta\cos\phi -x_i^\mathrm{toe} \right\}^2\right. \\
      &\left.\quad +\left\{y+(-1)^2(r+d)\cos\theta\cos\phi \right\}^2\right]^{\frac{1}{2}},
    \end{aligned}
    & \text{when Leg $i$ is in the stance phase}
  \end{cases}
\end{align*}
and $x^\mathrm{toe}_i$ is the contact position of Leg $i$.
Because Leg $i$ is at the nominal length $l_0$ during the swing phase, the leg spring force is generated only during stance.

From the measured cheetah data~\cite{Kamimura2021-ep}, we determined the physical parameters as follows: $m = 19$~kg, $J = 0.53$~kgm$^2$, $r = 0.29$~m, $l_0 = 0.69$~m, and $d = 0.06$~m. We used $k_0 = 100$~Nm/rad and $k = 15000$~N/m to reproduce a similar locomotor behavior to that of cheetahs from the perspective of body bending and gait cycle.

\subsection{Periodic solutions}
% モデルの周期解を求めるために，ポアンカレ断面を飛翔期の$\dot{y}=0$の瞬間に定義する．
% 両脚とも接地期を1回ずつ経験してから次のポアンカレ断面との交差が起こる解のみを探索する．
% 水平位置は走行中に単調に増加し，周期的ではないため，周期解を求める際には水平位置を無視する．
% 周期解は全て同一の力学的エネルギー$E=4500$ Jを持つように制限した．
% このエネルギーで発見される解はすべて，およそチーターと同程度の進行速度（18 m/sほど）を持つ．
% エネルギーを全ての解で同一にしているため，$y$, $\phi$, $\dot{\theta}$が決まると$\dot{x}$は一意に定められる．
% したがって，ポアンカレ断面上の固定点$z^*=[y^* \ \theta^* \phi^* \ \dot{\theta}^* \ \dot{\phi}^*]^\top$とそれに対応する接地角$u^*=[\gamma_1^\mathrm{td*} \ \gamma_2^\mathrm{td*}]^\top$をニュートン・ラフソン法によって数値的に探索し，周期解に対応する固定点を求める．
% ただし，探索する解は$\phi^*>0$のみに限定する（その理由は後述する）．

To find periodic solutions of the model, we defined the Poincar\'e section at the apex height of the COM ($\dot{y}=0$) during the flight phase.
We searched only for solutions in which both legs each experience one stance phase before the next intersection with the Poincar\'e section.
Because the horizontal position $x$ increases monotonically during running and is not periodic, we neglected the horizontal position when computing periodic solutions.
All periodic solutions were constrained to have the same mechanical energy, $E=4500$~J, where the solutions found had approximately the same horizontal velocity as cheetahs (about 18~m/s).
Because all solutions have the same total energy, once $y$, $\phi$, and $\dot{\theta}$ are given, $\dot{x}$ is uniquely determined.
We numerically searched for fixed points on the Poincar\'e section $z^*=[y^*\ \theta^*\ \phi^*\ \dot{\theta}^*\ \dot{\phi}^*]^\top$, and its corresponding touchdown angles $\gamma_1^\mathrm{td*}$ and $\gamma_2^\mathrm{td*}$ using the Newton-Raphson method.
We restricted the search to solutions with $\phi^*>0$ only (the reason is described later).

% 得られた周期解は，脚が接地する順番と，体幹の曲げ伸ばしの順番によって４種類に分類を行う(Fig.~\ref{solution_types})．
% 1周期のうちに2回のFlightを持つ解のうち，チーターと同様に，
% 肢離地後に体幹が伸びて前肢離地後に体幹が曲がる解，すなわち$\phi$が後肢離地後の飛翔期に負の極値、前肢離地後の飛翔期に正の極値をとる解をType EGとする．
% これに対して前後肢離地後に対する背骨の曲げ伸ばしが，Type EGと逆になる解をType GEとする．
% 2回のFlightでいずれも体幹が伸ばされる解をType EEとする．
% また，1周期のうちに1回しかFlightを持たず，flightで体幹が伸ばされる解をType Eとする．
% 探索範囲を$\phi^*<0$の範囲にも拡張すると，上記以外の解も発見されうるが，そのような範囲で見つかる解は，体幹運動が床反力を増加させ走行パフォーマンスが悪く，チーターの運動とは対応しないことが先行研究\cite{Kamimura2022-fa}からわかっている．

\begin{figure}
  \centering
  \includegraphics[scale = 1.0]{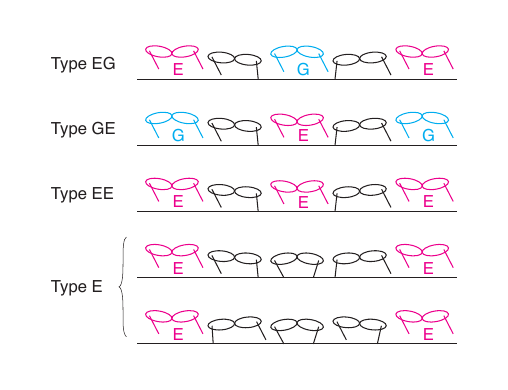}
  \caption{Classification of periodic solutions. Type EG: the spine is extended after the hindlimb stance, as in cheetahs. Type GE: the spine is extended after the forelimb stance. Type EE: the spine is extended after both the forelimb and hindlimb stance. Type E: the spine is extended during the flight.}
  \label{solution_types}
\end{figure}

We classified the obtained periodic solutions into four types according to the order of leg contact and the order of spinal flexion--extension (Fig.~\ref{solution_types}).
Among solutions with two flight phases in one period, we defined type EG as solutions in which, as in cheetahs, the spine is extended after hindlimb liftoff and flexed after forelimb liftoff; that is, $\phi$ takes a negative extremum in the flight phase after hindlimb liftoff and a positive extremum in the flight phase after forelimb liftoff.
In contrast, we defined type GE as solutions in which the spinal flexion--extension pattern after hindlimb and forelimb liftoff is the reverse of type EG.
We defined type EE as solutions in which the spine is extended in both flight phases.
We defined type E as solutions with only one flight phase in one period, in which the spine is extended during that flight phase.
If the search range is expanded to $\phi^*<0$, solutions other than the above can also be found; however, previous research~\cite{Kamimura2022-fa} has shown that such solutions do not correspond to cheetah locomotion because spinal motion increases the GRF and degrades running performance.

\subsection{Evaluation criteria}

% 発見された周期解に対して，最大床反力と平均進行速度，そして安定性を比較する．
% 最大床反力は，前後脚バネにかかる最大の力とする．
% 平均進行速度は，1ストライドにおける進行速度の時間平均とする．
% ポアンカレ断面上の固定点回りのポアンカレ写像のヤコビ行列を数値的に求め，その固有値がすべて1以下なら，その解は安定である．
% ただし系は保存系なので一つの固有値は必ず1であり，これは無視する．

We compared the obtained periodic solutions from the perspectives of maximum GRF, average horizontal velocity, and stability.
The maximum GRF is determined as the maximum force applied to the fore and hindlimb springs during running.
% なお，解の対称性からいずれの脚においても最大値は同じ値である．
Note that, due to the symmetry of the solutions, the maximum GRF is the same for both legs.
The average horizontal velocity $\bar{v}$ is defined as the time average of the horizontal velocity in one gait cycle.
% , as follows.
% \begin{align}
%   \bar{v}=\frac{1}{T}\int^T_0 v(t)\mathrm{d}t
% \end{align}
We numerically computed the Jacobian matrix of the Poincar\'{e} map around the fixed point on the Poincar\'{e} section.
If the absolute values of the eigenvalue are less than or equal to 1, the solution is stable.
Because the system is conservative, one eigenvalue is always 1, and this eigenvalue was ignored.

\section{RESULTS}
\subsection{Periodic solutions}
% 様々な$\kappa$について周期解を探索した結果，多くの解が得られた．
% $\kappa = \sqrt{10}$で発見された解のうち，チーターの走行と同様に2回の異なる飛翔期を含み，後肢接地のあとにExtended flightとなる解（EG）と，それに対してチーターと逆に前肢接地のあとにExtended flightとなる解（GE）の代表的なものの挙動をFig.~\ref{fig:vairables}に示す．
% Type EGの解は$[y^*, \dot{\theta}^*] = [0.67, -1.0]$で得られたもので，Type GEの解は$[y^*, \dot{\theta}^*] = [0.67, 1.0]$で得られたものである．
% いずれの解も$x(t), y(t), \phi(t)$の時間変化は定性的に類似している．
% これに対してtype EGの解では$\dot{\theta}(0)=\dot{\theta}^*$が負の値であるため，前肢(Leg~1)が後肢(Leg~2)よりも先に接地する．
% これに対してtype GEの解では$\dot{\theta}(0)=\dot{\theta}^*$が正の値であるため，後肢(Leg~2)が前肢(Leg~1)よりも先に接地する．
% すわなち，Type EGとGEでは，背骨の曲げ伸ばしを示す$\phi$の時間変化は同じなので，これを基準とすると，前後肢の接地タイミングと全体の姿勢角速度$\dot{\theta}$の正負のみが逆となる解が得られている。
% このような定性的な特徴は，$\kappa$の値にかかわらず，発見されたすべてのtype EGの解とtype GEの解でそれぞれ共通して見られる．

\begin{figure}
  \centering
  \includegraphics[width=\textwidth]{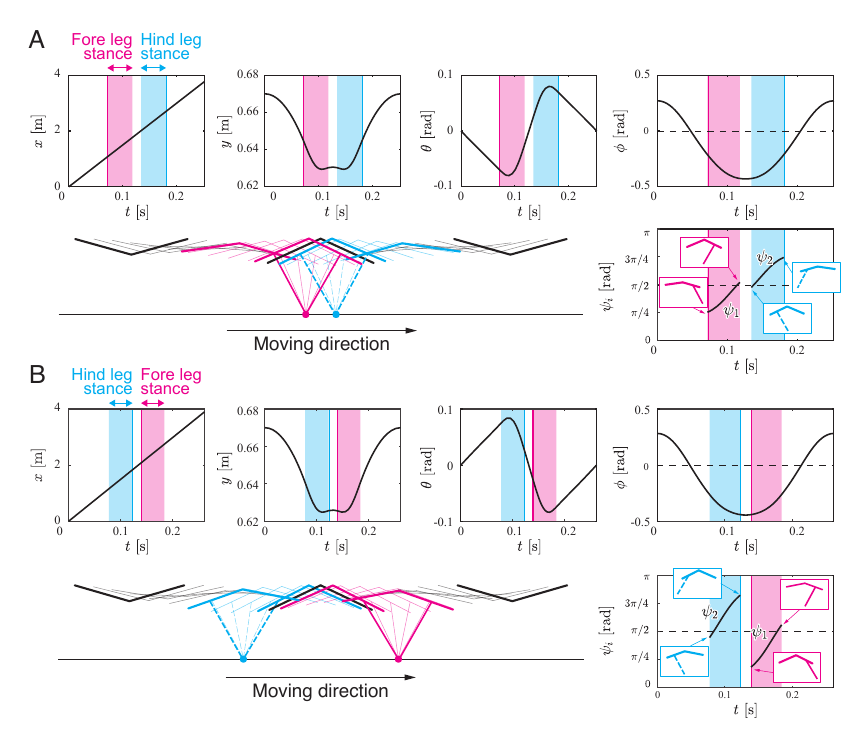}
  \caption{Time profiles of variables and relative leg angle $\psi_i$ ($i=1,2$) between the Body $i$ and the Leg $i$, and stick diagrams of typical solutions of (A) type EG ($[y^*, \dot{\theta}^*]=[0.67,-1.0]$, `a2' in Fig.~\ref{fig:sol_distri}) and (B) type GE ($[y^*, \dot{\theta}^*]=[0.67,1.0]$, `b2' in Fig.~\ref{fig:sol_distri}). Red and blue shaded areas indicate fore and hindlimb stance, respectively. Stick diagrams highlight the initial state, touchdown and liftoff of forelimb, mid-gait, touchdown and liftoff of hindlimb, and terminal state.}
  \label{fig:vairables}
\end{figure}

We found many periodic solutions over a range of $\kappa$ values.
Among the solutions found for $\kappa=\sqrt{10}$, Fig.~\ref{fig:vairables} shows two representative solutions with two distinct flight phases: type EG (Fig.~\ref{fig:vairables}A), which, like cheetah running, has extended flight after hindlimb liftoff, and type GE (Fig.~\ref{fig:vairables}B), which, unlike cheetah running, has extended flight after forelimb liftoff.
The type EG solution was obtained at $[y^*, \dot{\theta}^*] = [0.67, -1.0]$, and the type GE solution at $[y^*, \dot{\theta}^*] = [0.67, 1.0]$.
In both solutions, the time profiles of $x(t)$, $y(t)$, and $\phi(t)$ are qualitatively similar.
In contrast, in the type EG solution, $\dot{\theta}(0)=\dot{\theta}^*$ is negative, so the forelimb (Leg~1) contacts the ground before the hindlimb (Leg~2), whereas in the type GE solution, $\dot{\theta}(0)=\dot{\theta}^*$ is positive, so the hindlimb (Leg~2) contacts the ground before the forelimb (Leg~1).
Because the time profile of $\phi$ (spinal flexion--extension) is the same in type EG and type GE solutions, the difference in fore/hind touchdown timing is caused by the opposite signs of the whole-body angular velocity $\dot{\theta}$.
These qualitative features are common to all obtained type EG and type GE solutions, respectively, regardless of the value of $\kappa$.

% 接地期における体幹姿勢と脚バネの相対角度$\psi_i$は，type EGとtype GEでいずれも前肢後肢それぞれの接地期において単調に増加し（すなわち，脚が体幹に対して前方から後方に移動する），定量的にもそれほど違いがない．
% 具体的には、前肢では$\psi_i$は接地期前半に小さい値を取り（体幹と脚が平行に近くなり、脚は胴体の外側に伸展する），接地期後半に$\pi/2$に近くなる（体幹と脚が垂直に近くなり、脚は胴体の内側に屈曲する）．
% 後肢では$\psi_i$は接地期前半に接地期前半に$\pi/2$に近い値を取り（体幹と脚が垂直に近く，脚は胴体の内側に屈曲），接地期後半に$\pi$に近い値を取る（体幹と脚が平行に近くなり、脚は胴体の外側に伸展する）．
% しかしながら，Type EGとGEでは、背骨の曲げ伸ばしに対する前後肢の接地タイミングが逆なので，歩行周期全体でみると大きな違いを生み出す．
% 具体的には、歩行中期(50\% gait cycle)における背骨の曲げ$\phi$の極小値（最大曲げ）において、Type EGでは$\psi_i$が$\pi/2$に近く、Type GEでは0に近くなる。

During stance, the relative angle $\psi_i$ between Body $i$ and the Leg $i$ increases monotonically in both type EG and type GE, for both forelimb and hindlimb stances (i.e., the leg moves from an anterior to a posterior position relative to the trunk), with only small quantitative differences.
Specifically, during forelimb stance, $\psi_i$ is small in the first half of stance (the body and leg are nearly parallel, and the leg is extended outward relative to the trunk) and approaches $\pi/2$ in the second half (the body and leg are nearly perpendicular, and the leg is flexed inward relative to the trunk).
During hindlimb stance, $\psi_i$ is close to $\pi/2$ in the first half of stance (the body and leg are nearly perpendicular) and approaches $\pi$ in the second half (the body and leg become nearly parallel).
However, because the forelimb/hindlimb contact timing relative to spinal flexion--extension is reversed between type EG and type GE, this leads to a substantial difference over the entire gait cycle.
Specifically, at the minimum of spinal bending $\phi$ (maximum flexion) at mid-gait (50\% gait cycle), $\psi_i$ is close to $\pi/2$ in type EG, whereas it is close to $0$ or $\pi$ in type GE.

% Fig.~\ref{fig:grf_v}で，Fig.~\ref{fig:vairables}に示したtype EGの解とtype GEの解における床反力と水平速度の時間変化を比較する．
% いずれも接地期前半に脚から床反力を受けて減速し，接地期後半に加速に転ずる．
% type EGの解では，type GEの解と比較して受ける床反力が小さいため，水平方向の減速量と加速量が小さい．
% そのため，平均進行速度はtype EGのほうが大きい．

\begin{figure}
  \centering
  \includegraphics{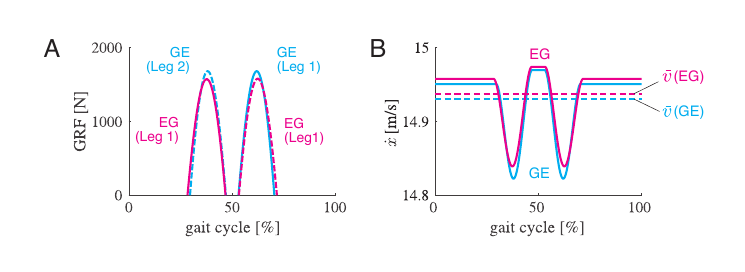}
  \caption{Time profiles of (A) GRF and (B) horizontal velocity of solutions with $[y^*, \dot{\theta}^*]=[0.67, -1.0]$ (type EG) and $[0.67, 1.0]$ (type GE). Dashed lines in (B) indicate average horizontal velocity.}
  \label{fig:grf_v}
\end{figure}

Figure~\ref{fig:grf_v} compares the time profiles of GRF and horizontal velocity between the type EG and type GE solutions shown in Fig.~\ref{fig:vairables}.
In both solutions, the model decelerates in the first half of stance due to GRF from the leg and then transitions to acceleration in the second half of stance.
In the type EG solution, the GRF is smaller than in the type GE solution, resulting in smaller horizontal deceleration and acceleration.
Therefore, the average horizontal velocity is higher in type EG.

\begin{figure}
  \centering
  \includegraphics[width=\linewidth]{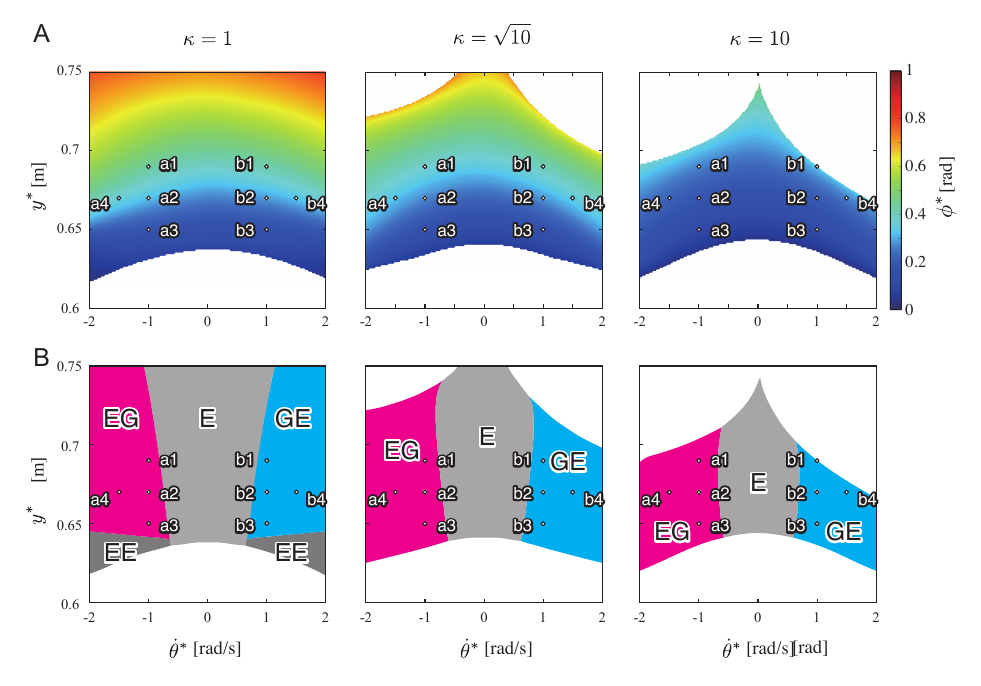}
  \caption{Distributions of fixed points on a Poincar\'{e} section in $y^*$-$\dot{\theta}^*$ plane for  $\kappa =1$ (left), $\sqrt{10}$ (middle), and $10$ (right). No solution was found in the white area. (A) Contour map of $\phi^*$. (B) Classification of solutions. Colors indicate types of solutions: EG (red), GE (blue), E (light grey), and EE (dark grey). White dots indicate the typical solutions of type EG (a1, a2, a3, and a4 indicate $[y^*, \dot{\theta}^*]=[0.69, -1.0]$, $[0.67, -1.0]$, $[0.65, -1.0]$, and $[0.67, -1.5]$, respectively) and type GE (b1,b2, b3, and b4 indicate $[y^*, \dot{\theta}^*]=[0.69, 1.0]$, $[0.67, 1.0]$, $[0.65, 1.0]$, and $[0.67, 1.5]$, respectively) used for the evaluation of periodic solutions.}
  \label{fig:sol_distri}
\end{figure}

% 得られた解は全て，Fig.~\ref{fig:vairables}に示した解と同様に，時間的に対称で(scissor symmetry~\cite{Poulakakis2006-ww})，$\theta^*=\dot{\phi}^*=0$が満たされていた．
% 以上から，発見された解を$y^*$-$\phi^*$-$\dot{\theta}^*$空間にプロットすると，$\kappa$の値に依存する，一つの滑らかな曲面上に分布していた．
% 代表的な解分布として，$\kappa =1, \sqrt{10}, 10$において$y^*$-$\dot{\theta}^*$平面に射影した解(この平面では一意に存在，ただし解が存在しないところも存在)とその分類をFig.~\ref{fig:sol_distri}に示す．
% $|\dot{\theta}^*|$があまり大きくないときにtype Eの解が得られた．
% $|\dot{\theta}^*|$が十分に大きいときに，1周期のうちに2回の異なる飛翔期を含む解が得られ，$\dot{\theta}^*>0$のときにtype EG，$\dot{\theta}^*<0$のときにtype GEと分類された．
% $\kappa$を大きくするにつれて，解が発見される$\dot{\theta}^*$の範囲はあまり変わらなかったのに対して，$y^*$と$\phi^*$の範囲が小さくなっていった．
% 特にtype GEの解が存在する$y^*$の範囲はtype EGの解が存在する範囲よりも狭い．
% 2回の飛翔期を持つが，いずれも体幹が伸ばされるType EEの解は$\kappa$が小さい場合にしか見つからなかった．

All obtained solutions, including those shown in Fig.~\ref{fig:vairables}, were temporally scissor symmetric \cite{Raibert1986-uv,Poulakakis2006-ww,Kamimura2022-fa} and satisfied $\theta^*=\dot{\phi}^*=0$.
Therefore, when plotted in the $y^*$-$\phi^*$-$\dot{\theta}^*$ space, the discovered solutions lay on a single smooth surface that depended on $\kappa$.
As representative examples, Fig.~\ref{fig:sol_distri} shows the solutions projected onto the $y^*$-$\dot{\theta}^*$ plane (where each solution is uniquely represented, although no solutions were found in some regions) and their classification for $\kappa =1$, $\sqrt{10}$, and $10$.
Type E solutions were obtained when $|\dot{\theta}^*|$ was relatively small.
When $|\dot{\theta}^*|$ was sufficiently large, solutions with two distinct flight phases per cycle were obtained and classified as type EG for $\dot{\theta}^*>0$ and type GE for $\dot{\theta}^*<0$.
As $\kappa$ increased, the range of $\dot{\theta}^*$ over which solutions were found changed little, whereas the ranges of $y^*$ and $\phi^*$ became narrower.
In particular, the $y^*$ range for type GE solutions was narrower than that for type EG solutions.
Type EE solutions, which have two flight phases with trunk extension in both, were found only for small $\kappa$.

% 前述のように，背骨の曲げ伸ばしに対する前後肢の接地タイミングが逆の関係にあるType EGとGEでは，歩行中期における体幹に対する脚の相対角度$\psi_i$に違いが見られた．
% この$\psi_i$はFig.~\ref{fig:phipsi}Aに示すように，$\psi_i$はそれぞれの脚$i$で接地した瞬間に最小となり，離地する瞬間に最大となる．
% また，一周期の中で最も相対角度が小さくなるのは，解の種類にかかわらず前肢(Leg~1)が接地する瞬間の$\psi_1^\mathrm{td}$で，相対角度が最も大きくなるのは後肢(Leg~2)が離地する瞬間の$\psi_2^\mathrm{td}$である．
% Figs.~\ref{fig:phipsi}B and Cに，特にtype EGとGEの解のみに着目して，$\kappa=1$, $\kappa=\sqrt{10}$, $\kappa=10$における$y^*$-$\dot{\theta}^*$平面内のLeg~1が接地する瞬間における体幹の曲げ角$\phi^\mathrm{td,1}$と脚バネのなす相対角度$\psi_1^\mathrm{td}$の$y^*$-$\dot{\theta}^*$平面内の分布と$\kappa$依存性を示す．
% まず同一の$\kappa$で比較すると，type EGの解の方が，type GEの解と比較して$\phi^\mathrm{td,1}$が0に近く（曲げが小さい），$\psi_1^\mathrm{td}$が$\pi/2$に近い傾向がある．
% 次に$\kappa$を変化させると，$\kappa$が大きくなるにつれて解が発見される$y^*$の上限が小さくなっていくことから，$y^*$の値が大きい点a1やb1の解において，$\phi^\mathrm{td,1}$と$\psi_1^\mathrm{td}$は小さくなる．
% Fig.~\ref{fig:phipsi}Dに，$\kappa$の値に対してtypeごと（type EGとGEのみ）に発見された全ての解における$\phi^\mathrm{td,1}$と$\psi_1^\mathrm{td}$の平均値と標準偏差を算出した結果を示す．
% $\kappa$の値にかかわらず，type EGの解の方がtype GEの解と比較して$\phi^\mathrm{td,1}$が0に近く（曲げが小さい），$\psi_1^\mathrm{td}$が\textcolor{red}{$\pi/2$に近い}傾向がある．
% $\kappa=2$程度までは，$\kappa$の値が大きくなるにつれて$\phi^\mathrm{td,1}$も$\psi_1^\mathrm{td}$も平均値が小さくなるが，それ以上の$\kappa$の値では平均値・標準偏差ともに大きな変化は見られない．
% なお，解は50\% Gait cycleに対して対称であることから，$\psi_2^\mathrm{lo} = \pi-\psi_1^\mathrm{td}$が成り立つ．そのため，後肢の離地する瞬間の相対角度$\psi_2^\mathrm{lo}$についても同様で，type EGの解の方がtype GEの解と比較して$\psi_2^\mathrm{lo}$が$\pi/2$に近い傾向がある．

\begin{figure}
  \centering
  \includegraphics[width=\textwidth]{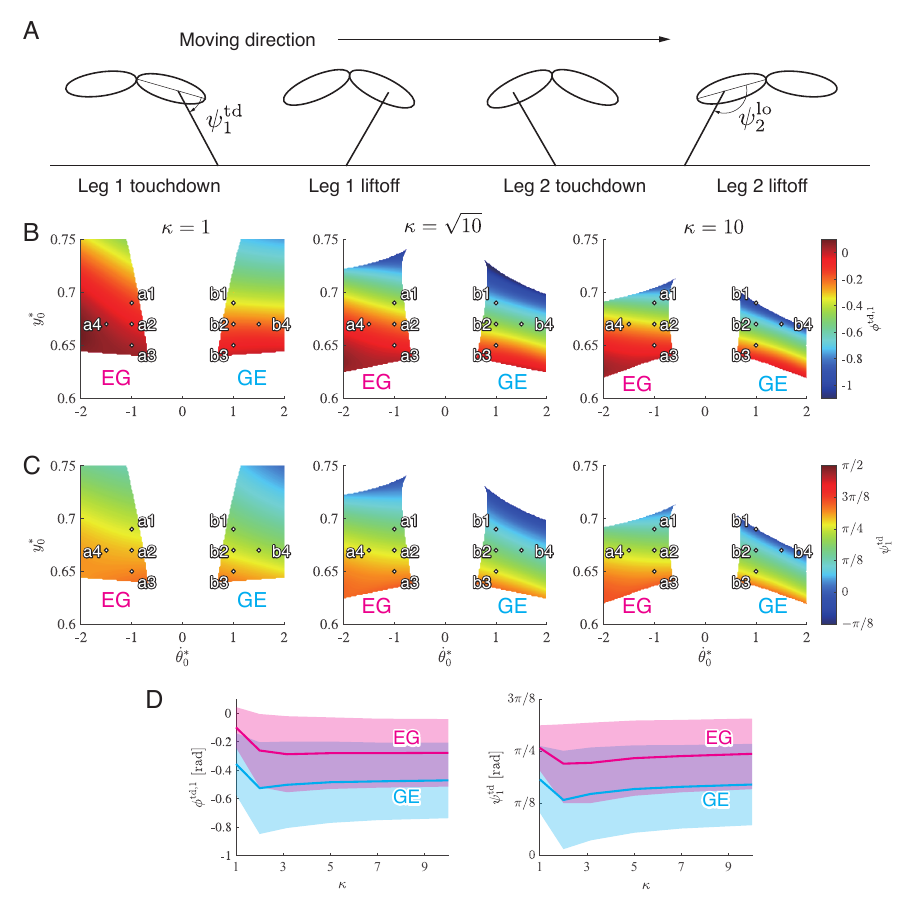}
  \caption{%
    % (A) 脚と体幹の相対角度$\psi_i$と接地・離地イベントの関係．$\kappa=1,\sqrt{10},10$において発見された全てのEG・GE解における，Leg~1が接地する瞬間の(B) 体幹の相対角度$\phi$と (C)脚バネのなす相対角度$\psi_1^\mathrm{td}$．(D) $\phi^\mathrm{td,1}$と$\psi_1^\mathrm{td}$の解のtypeごとの平均値と標準偏差の$\kappa$依存性．
    Relationship between the leg--body relative angle $\psi_i$ and spine motion.
    (A) Relationship between $\psi_i$ and touchdown/liftoff events.
    (B) Spine bending angle $\phi$ and (C) leg--trunk relative angle $\psi_1^\mathrm{td}$ at Leg~1 touchdown for all obtained types EG and GE solutions at $\kappa=1$, $\sqrt{10}$, and $10$.
    (D) $\kappa$ dependence of the mean and standard deviation of $\phi^\mathrm{td,1}$ and $\psi_1^\mathrm{td}$ for each solution type.
  }
  \label{fig:phipsi}
\end{figure}

As described above, in types EG and GE, where forelimb and hindlimb contact timings are opposite relative to spinal flexion--extension, differences appear in the leg--trunk relative angle $\psi_i$ during mid-stance.
% The GRF is smaller when $\psi_i$ is closer to $\pi/2$ and larger when it is closer to $0$ or $\pi$ (Fig.~\ref{fig:grf_v}).
As shown in Fig.~\ref{fig:phipsi}A, for each Leg $i$, $\psi_i$ takes its minimum at touchdown and its maximum at liftoff.
Moreover, within one gait cycle, the smallest relative angle is $\psi_1^\mathrm{td}$ at forelimb (Leg~1) touchdown, and the largest is $\psi_2^\mathrm{lo}$ at hindlimb (Leg~2) liftoff, regardless of solution type.
Figs.~\ref{fig:phipsi}B and C focus on type EG and GE solutions and show, for $\kappa=1$, $\sqrt{10}$, and $10$, the distributions on the $y^*$-$\dot{\theta}^*$ plane and the $\kappa$ dependence of the spine bending angle at Leg~1 touchdown $\phi^\mathrm{td,1}$ and the leg--body relative angle $\psi_1^\mathrm{td}$.
For a single $\kappa$, compared with type GE solutions, type EG solutions tend to have $\phi^\mathrm{td,1}$ values closer to $0$ (smaller flexion) and $\psi_1^\mathrm{td}$ values closer to $\pi/2$.
As $\kappa$ increases, the upper bound of $y^*$ for discovered solutions decreases; therefore, for solutions such as points `a1' and `b1' with larger $y^*$, both $\phi^\mathrm{td,1}$ and $\psi_1^\mathrm{td}$ decrease.
Fig.~\ref{fig:phipsi}D shows the mean and standard deviation of $\phi^\mathrm{td,1}$ and $\psi_1^\mathrm{td}$ over all obtained solutions for each type (types EG and GE only) across values of $\kappa$.
Regardless of $\kappa$, compared with type GE solutions, type EG solutions tend to have $\phi^\mathrm{td,1}$ values closer to $0$ (smaller flexion) and $\psi_1^\mathrm{td}$ approach to $\pi/2$.
Up to approximately $\kappa=2$, the mean values of both $\phi^\mathrm{td,1}$ and $\psi_1^\mathrm{td}$ decrease as $\kappa$ increases; for larger $\kappa$, neither the mean values nor the standard deviations change substantially.

\subsection{Gait performance}
% $\kappa=1$, $\sqrt{10}$, and $10$のそれぞれにおいて，得られたtypes EG and GEの周期解の最大床反力，平均進行速度，最大固有値を評価した結果をそれぞれFigs.~\ref{fig:criteria_distribution}A, B, and Cに示す．
% $\kappa$の値に関わらず，$y^*$が大きいほど最大床反力が小さく，それに伴い平均進行速度が小さくなる．ただし，最大床反力の減少割合に対して平均進行速度の減少割合は小さい．
% また，$\kappa$が大きくなると最大固有値は大きくなり，安定な解の領域が狭くなっていく．
% また，$\kappa$の値に対する依存性を調べるため，発見されたtypes EG and GEの解すべてに対して最大床反力と平均進行速度を$\kappa$ごとに平均と標準偏差を算出した結果をFig.~\ref{fig:criteria_distribution}Dに，$y^*$-$\dot{\theta}^*$平面で安定な解が存在する領域の面積をFig.~\ref{fig:criteria_distribution}Eに示す．
% $\kappa$の値を大きくすると，いずれの種類の解も最大床反力は小さくなるのに対して，平均進行速度に大きな変化は見られない．
% また，type EGは最大床反力の平均値がtype GEの平均値よりも小さく，平均進行速度の平均値がtype GEの平均値よりもわずかに小さい．
% 最大固有値については，$\kappa$の値が大きくなるにつれて，安定な解の領域が狭くなっていく．特にtype GEは$\kappa=1$付近ではtype EGよりも安定領域が広いものの，$\kappa$が大きくなるにつれてその領域は急激に狭くなっていく．

\begin{figure}
  \centering
  \includegraphics[width=\textwidth]{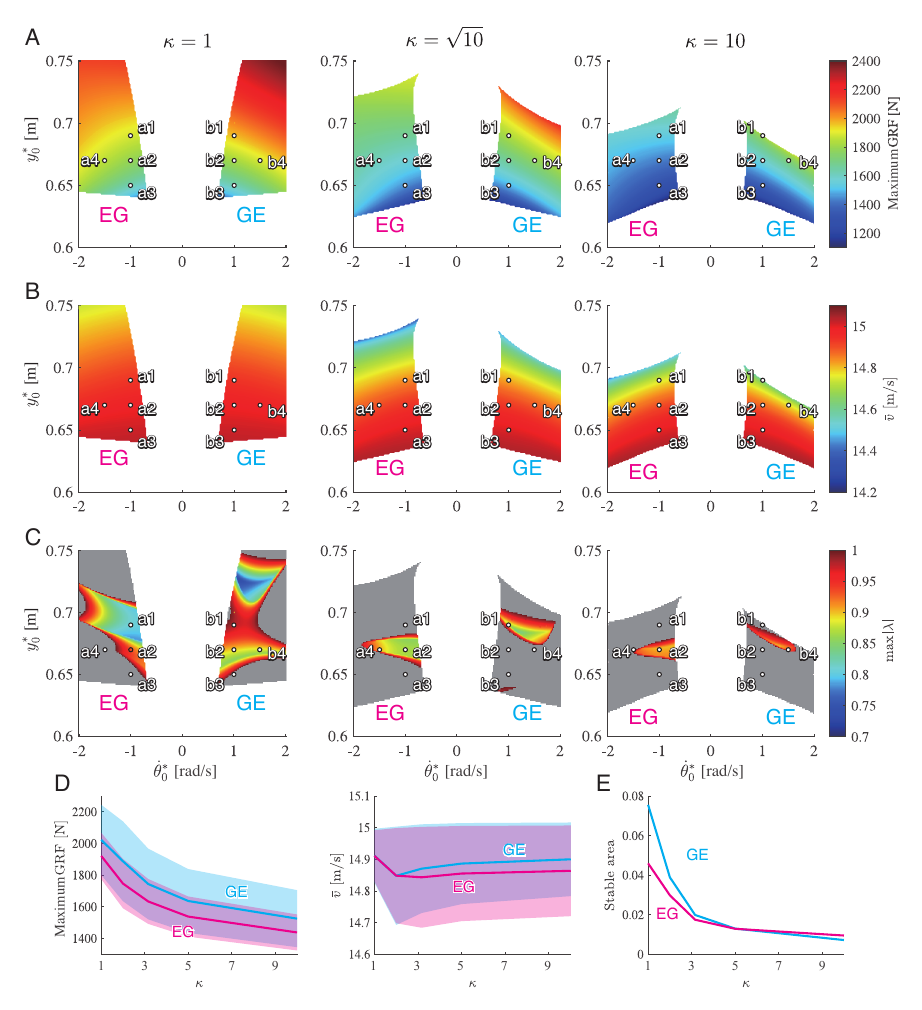}
  \caption{Evaluation criteria of solutions of types EG and GE for $\kappa=1$ (left), $\sqrt{10}$ (center), and $10$ (right). White lines indicate the boundaries of solution types. (A) Maximum GRF. (B) Average horizontal velocity. (C) Maximum absolute eigenvalue. The gray area indicates the region where solutions are unstable. (D) Performance dependence of solutions of types EG and GE on $\kappa$. Solid lines and shaded areas indicate the mean and standard deviation of the solutions of types EG and GE, respectively.}
  \label{fig:criteria_distribution}
\end{figure}

For $\kappa=1$, $\sqrt{10}$, and $10$, Figs.~\ref{fig:criteria_distribution}A, B, and C show the maximum GRF, average horizontal velocity, and maximum eigenvalue of the obtained periodic solutions of types EG and GE, respectively.
Regardless of the value of $\kappa$, a larger $y^*$ corresponds to a smaller maximum GRF and, accordingly, a smaller average horizontal velocity.
However, the reduction rate in average horizontal velocity is smaller than that in maximum GRF.
In addition, as $\kappa$ increases, the maximum eigenvalue increases and the region of stable solutions becomes smaller.
To further investigate the dependence on $\kappa$, for all obtained solutions of types EG and GE, we calculated the mean and standard deviation of the maximum GRF and average horizontal velocity for each $\kappa$; the results are shown in Fig.~\ref{fig:criteria_distribution}D.
The area of the region where stable solutions exist on the $y^*$-$\dot{\theta}^*$ plane is shown in Fig.~\ref{fig:criteria_distribution}E.
As $\kappa$ increases, the maximum GRF decreases for both types of solution, whereas no significant change is observed in average horizontal velocity.
Moreover, type EG has a smaller mean maximum GRF than type GE and a slightly smaller mean average horizontal velocity than type GE.
The stable-solution region becomes smaller as $\kappa$ increases.
In particular, type GE has a wider stable region than type EG around $\kappa=1$, but this region shrinks rapidly as $\kappa$ increases.

% Fig.~\ref{fig:kappa_dependency}はFig.~\ref{fig:sol_distri}に示す8つの代表解a1, a2, a3, a4（type EG）とb1, b2, b3, b4（type GE）について，様々な$\kappa$に対する安定性，最大床反力，平均水平速度を示す．
% いずれのtypeの解でも，$\kappa$が大きくなるにつれて最大床反力と平均水平速度は小さくなる．
% $\kappa$の値にかかわらず，type EGの解の最大床反力はtype GEの解より小さい．
% $\kappa$が十分大きい場合（およそ$\kappa>2$）には，差は小さいもののtype GEの解の平均水平速度がtype EGの解より大きい．
% $\kappa$が大きくなると，type EGの解は徐々に安定性を失って最終的に不安定になり，これに対してtype GEの解は急速に安定性を失う．

\begin{figure}
  \centering
  \includegraphics[width=\textwidth]{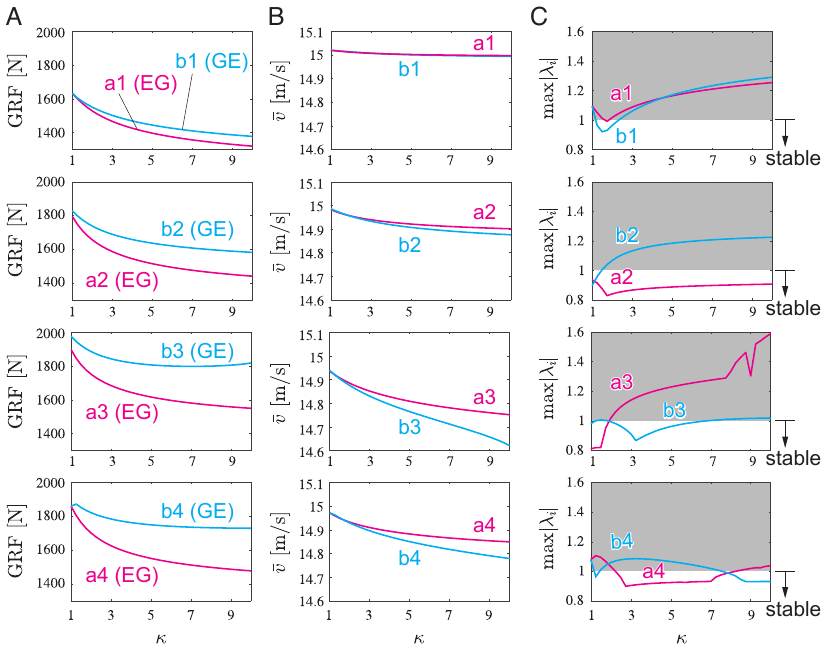}
  \caption{Evaluation criteria for various $\kappa$ for typical solutions of types EG (a1, a2, a3, and a4 in Fig.~\ref{fig:sol_distri}) and GE (b1, b2, b3, and b4 in Fig.~\ref{fig:sol_distri}). (A) Maximum GRF. (B) Average horizontal velocity. (C) Maximum eigenvalue. The gray area indicates the region where the solution is unstable.}
  \label{fig:kappa_dependency}
\end{figure}

Figure~\ref{fig:kappa_dependency} shows stability, maximum GRF, and average horizontal velocity for the eight representative solutions a1, a2, a3, a4 (type EG) and b1, b2, b3, b4 (type GE) shown in Fig.~\ref{fig:sol_distri} across various $\kappa$ values.
For both types of solution, the maximum GRF and the average horizontal velocity decreased as $\kappa$ increased.
Regardless of the value of $\kappa$, the type EG solution has a smaller maximum GRF than the type GE solution.
If $\kappa$ is sufficiently large (approximately $\kappa>2$), the solution of type GE has a higher average horizontal velocity than that of type EG, although the differences are small.
As $\kappa$ increased, the solution of type EG gradually lost stability and eventually became unstable.
In contrast, the solution of type GE rapidly lost stability as $\kappa$ increased.

\section{DISCUSSION}
\subsection{Difference between types EG and GE solutions}
% 本研究では，体幹の曲げ伸ばしに対する前後肢の接地順の関係が走行に与える影響を，シンプルな動力学モデルを用いて調べた．
% 先行研究\cite{Kamimura2022-fa}では，体幹の曲げ伸ばし運動は飛翔期よりもむしろ接地期に重要な役割を果たしており，接地時に体幹バネが適切な姿勢になっていれば，走行中の床反力が低減されることが示された．
% 接地期において体幹が山形(`$\wedge$'-shape)に曲げられている場合，体幹バネが発するトルクは，脚の付け根に上向きの力を発することと等価である．
% すなわち，脚の付根を持ち上げる効果をもたらす．
% このとき，脚の付け根で体幹バネが発する力と床反力はいずれも上向きになるため，体幹バネが床反力を低減させる(Fig.~\ref{fig:grf_mechanism}a)．
% これによって，接地期前半における減速量が小さくなることで，平均進行速度が大きく保たれる(Fig.~\ref{fig:grf_v})．
% このような運動は，$\phi^*>0$となる解で達成されることが先行研究~\cite{Kamimura2022-fa}で示されている．
% そこで，本研究ではそのような解に限って探索を行った(Fig.~\ref{fig:sol_distri})．
% 発見された1周期の間に2回の異なる飛翔期を持つ解(type EGとGE)は全て，Fig.~\ref{fig:vairables}に示した解と同様に，接地期において常に$\phi<0$を満たし，体幹が`$\wedge$'形に曲げられているものであった．
% しかしながら，床反力が適切に低減されるのは体幹が適度に曲がられていて，体幹部と床反力の向きが垂直に近い場合（Fig.~\ref{fig:grf_mechanism}a）である．
% これに対して，体幹の曲げが過度に大きく体幹と脚が平行に近くなると，体幹バネが足の付根に発生させる力は床反力と垂直に近くなり，床反力を低減させる効果は小さくなってしまう（Fig.~\ref{fig:grf_mechanism}b）．

\begin{figure}
  \centering
  \includegraphics[scale=1.0]{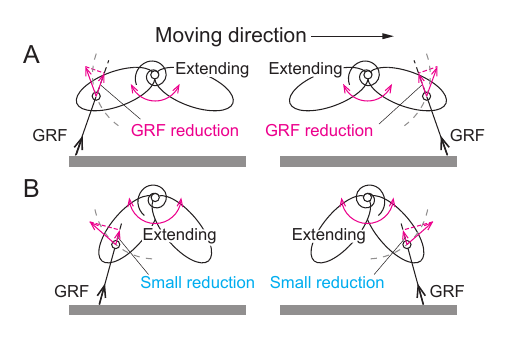}
  \caption{Relationship between GRF and body spring torque. (a) The spine is moderately flexed. (b) The spine is excessively flexed. (Left) hindlimb stance. (Right) forelimb stance.}
  \label{fig:grf_mechanism}
\end{figure}

In this study, using a simple model, we investigated how running performance of cheetahs is affected by the forelimb--hindlimb contact sequence relative to spinal flexion--extension.
Our previous study~\cite{Kamimura2022-fa} showed that spinal flexion--extension plays a more important role during stance than during flight, and that the GRF can be reduced when the spine takes an appropriate posture during stance.
Specifically, when the spine is flexed into a wedge shape (`$\wedge$'-shape) during stance, the torque generated by the body spring is equivalent to generating an upward force at the leg root.
In other words, it provides a lifting effect at the leg root.
In this case, because both the force generated by the body spring at the leg root and the GRF are directed upward, the body spring reduces the GRF (Fig.~\ref{fig:grf_mechanism}A).
As a result, deceleration in the first half of stance is reduced, thereby maintaining a higher average horizontal velocity (Fig.~\ref{fig:grf_v}).
Our previous study~\cite{Kamimura2022-fa} also showed that such motion is achieved by solutions with $\phi^*>0$.
Therefore, in this study, we searched only for such solutions (Fig.~\ref{fig:sol_distri}).
All discovered solutions with two distinct flight phases within one gait cycle (types EG and GE), like the solution shown in Fig.~\ref{fig:vairables}, satisfied $\phi<0$ throughout stance and had the spine flexed into a `$\wedge$'-shape.
However, the GRF is appropriately reduced only when the spine is moderately flexed and the trunk orientation is nearly perpendicular to the GRF direction (Fig.~\ref{fig:grf_mechanism}A).
In contrast, when spinal flexion is excessively large and the trunk becomes nearly parallel to the leg, the force generated by the body spring at the leg root becomes nearly perpendicular to the GRF, and the GRF-reduction effect becomes small (Fig.~\ref{fig:grf_mechanism}B).

% 本研究では，前肢と後肢の接地順序そのものではなく，前後肢の接地期が背骨の曲げ・伸ばし運動のどの位相に位置付けられているかという相対的な位相関係に着目し，1周期に2回の異なる飛翔期を持つ2種類の解を，後肢接地期後にチーターと同様に背骨を伸ばすtype EGと，チーターと違って背骨を曲げるtype GEに分類した．
% 得られた結果から，接地と背骨運動の相対的な位相関係が，以下のようなメカニズムで脚にかかる床反力を低減し，走行のパフォーマンスに大きな影響を与えることが示される．
% 脚は前後いずれも接地した瞬間に体幹に対して最も前方に位置し，接地期を通じて体幹に対して後方に移動し，離地する瞬間に最も後方に位置する．
% これに対して，接地期において体幹曲げ角$\phi$は常に負の値を取り，その時間変化は脚の角度変化と比較して大きくない．
% このため，解の種類にかかわらず，前肢(Leg~1)接地期の始まりと後肢(Leg~2)接地期の終わりにおいて，体幹と脚が平行に近くなる(Fig.~\ref{fig:phipsi}A)．
% これらの期間においては，相対角度が$\pi/2$に近くなる前肢の接地期後半や後肢の接地期前半と比較して，体幹バネが床反力を減少させる効果は小さい．
% いずれのタイプの解においても，$\phi$の振る舞いは定性的に同じであり，体幹曲げは50\% gait cycleで極値を取り，最も曲げられる．
% ただし，type EGでは$\phi$が極値から遠く，それほど体幹が曲げられていない期間に，前肢接地の始まりと後肢接地の終わりが位置している．
% これに対してtype GEでは$\phi$が極値に近く，大きく体幹が曲げられている期間に，前肢接地の始まりと後肢接地の終わりが位置している．
% したがって，type GEでは，より体幹が曲げられた時点で脚が体幹と平行に近くなるタイミングを迎えるため，type EGよりもさらに体幹と脚は平行に近づき(Fig.~\ref{fig:phipsi}C and D)，床反力の低減効果が小さくなってしまう．
% したがって，$\kappa$の値にかかわらず，type EGの解はtype GEの解よりも床反力低減効果が大きく，最大床反力が小さく抑えられる傾向にある(Figs.~\ref{fig:criteria_distribution}D and~\ref{fig:kappa_dependency})．
% さらに，$y^*$が大きくない範囲の解においては，$\kappa$の値にかかわらず，type EGの解はtype GEの解よりも平均進行速度が大きい傾向にある(Fig.~\ref{fig:kappa_dependency})．

In this study, we focused not on the forelimb--hindlimb contact order itself, but on the relative phase relationship between stance phases and spinal flexion--extension motion.
Accordingly, we classified the two types of solutions with two distinct flight phases per gait cycle into type EG, in which the spine is extended after hindlimb stance as in cheetahs, and type GE, in which the spine is flexed after hindlimb stance unlike in cheetahs.
The obtained results indicate that, rather than the contact order itself, the relative phase relationship between contact and spinal motion strongly affects running performance by reducing the GRF through the following mechanism.
At touchdown, both forelimb and hindlimb are positioned most anteriorly with respect to the fore and hind bodies; during stance, they move posteriorly relative to the trunk, and at liftoff, they are positioned most posteriorly.
In contrast, during stance, the spinal flexion angle $\phi$ is always negative in the discovered solutions, and its temporal variation is not large compared with the angular change of the legs.
Therefore, regardless of solution type, the body and leg become nearly parallel at the beginning of forelimb (Leg~1) stance and at the end of hindlimb (Leg~2) stance (Fig.~\ref{fig:phipsi}A).
As a result, during these periods, the body spring has a smaller GRF-reduction effect than in the latter half of forelimb stance and the first half of hindlimb stance, where the relative angle is closer to $\pi/2$.
In both types, the qualitative behavior of $\phi$ is the same: it takes an extremum at the 50~\% gait cycle, where spinal flexion is maximal.
However, in type EG solutions, the beginning of forelimb stance and the end of hindlimb stance occur when $\phi$ is still far from this extremum and spinal flexion is not very large, whereas in type GE solutions these events occur when $\phi$ is close to the extremum and spinal flexion is large.
Consequently, in type GE solutions, the timing at which the leg becomes nearly parallel to the body occurs at a more strongly flexed spinal posture; therefore, the body and leg become even closer to parallel than in type EG solutions (Fig.~\ref{fig:phipsi}C and D), and the GRF-reduction effect becomes smaller.
Therefore, regardless of the value of $\kappa$, type EG solutions tend to exhibit a larger GRF-reduction effect and a smaller maximum GRF than type GE solutions (Figs.~\ref{fig:criteria_distribution}D and~\ref{fig:kappa_dependency}).
Furthermore, in the range of solutions with relatively small $y^*$, type EG solutions tend to exhibit higher average horizontal velocity than type GE solutions, regardless of the value of $\kappa$ (Fig.~\ref{fig:kappa_dependency}).

\subsection{The effect of asymmetric body flexibility}

% 発見されたtype EGとGEの解は，いずれも$\kappa$の値を大きくするほど，床反力が小さくなるのに対して，平均進行速度はそれほど変化しない傾向がある(Figs.~\ref{fig:criteria_distribution}D and~\ref{fig:kappa_dependency})．
% このため，$\kappa$を大きくすることで，平均進行速度を保ちつつ脚にかかる負担を減らすことができる．
% しかし，$\kappa$を大きくするほど，安定な解が存在する領域が狭くなっていくため（Fig.~\ref{fig:criteria_distribution}E），$\kappa$を大きくすることには限界がある．
% このトレードオフ関係のため，高い安定性を保ちながら床反力を小さくするような最適な$\kappa$は存在し得ない．
% なお，$\kappa$の増加に対して床反力と安定領域は，$\kappa$が小さい範囲($\kappa=3$程度まで)で大きく変化し，それ以上に大きな$\kappa$ではほとんど変化しない．

For both discovered solution types with two distinct flight phases (EG and GE), GRF decreases as $\kappa$ increases, whereas average horizontal velocity changes little (Figs.~\ref{fig:criteria_distribution}D and~\ref{fig:kappa_dependency}).
Therefore, increasing $\kappa$ can reduce leg loading while maintaining average horizontal velocity.
However, as $\kappa$ increases, the region in which stable solutions exist becomes smaller (Fig.~\ref{fig:criteria_distribution}E); thus, there is a practical limit to how large $\kappa$ can be.
Because of this trade-off, no single $\kappa$ simultaneously achieves high stability and small GRF.
Note that, the reductions in GRF and stable-region area are pronounced when $\kappa$ is relatively small (up to approximately $\kappa=3$), but they remain almost unchanged for larger values of $\kappa$.

% また，type EGの解はtype GEの解と比較して，$\kappa$の値にかかわらず，床反力が小さい．
% さらに，より高速な走行が実現できる$y^*$が小さい領域で比較すると(Fig.~\ref{fig:kappa_dependency})，type EGの解はtype GEの解よりも平均進行速度が大きい．
% 上記の傾向は，$\kappa$が小さく1に近いとそれほど大きな差ではないが，$\kappa$を大きくするほど顕著になる．

Compared with type GE solutions, type EG solutions exhibit smaller GRF regardless of the value of $\kappa$.
Moreover, in the small-$y^*$ region, where higher-speed running can be achieved (Fig.~\ref{fig:kappa_dependency}), type EG solutions have higher average horizontal velocities than type GE solutions.
This tendency is weak when $\kappa$ is small and close to 1, but becomes increasingly evident as $\kappa$ increases.

% 上記から，脚の負担を大きくせずに安定な高速走行を実現するためには，体幹剛性の非対称性$\kappa$を大きくしすぎないことに加えて，type EGのような体幹曲げと脚の向きの関係を持つ歩容を用いることが重要であることが示される．

These results indicate that, to achieve stable high-speed running without increasing leg loading, it is important not only to avoid making the spinal-stiffness asymmetry $\kappa$ excessively large but also to use a gait with an appropriate phase relationship between spinal motion and forelimb--hindlimb stance, as in type EG.

\subsection{Cheetah running and spine movement}

% 先行研究~\cite{Kamimura2022-fa}では，EG解とGE解の走行性能に安定性以外の指標では大きな差異は見られなかった．
% しかし，本研究の結果から，チーターのようにEG解に相当する歩容は，GE解に相当する歩容と比較して，体幹曲げと脚の向きの関係から，本質的に床反力が小さくなることが明らかになった．
% また，高速走行を実現する小さな$y^*$の領域では，EG解はGE解よりも平均進行速度が大きい傾向がある．
% これらのことから，チーターはtype EGのような背骨運動と接地期の位相関係を用いることで，背骨の曲げ運動を効果的に用いて脚の負担を減らしつつ高速走行を実現していることが示唆される．

In our previous study~\cite{Kamimura2022-fa}, no substantial differences were found between types EG and GE solutions in running-performance metrics other than stability.
However, our current results revealed that a gait corresponding to the type EG solution, as in cheetahs, intrinsically yields smaller GRF than a gait corresponding to the GE solution because of differences in the relative body--leg angle.
In addition, in the small-$y^*$ region that enables high-speed running, type EG solutions tend to show higher average horizontal velocity than GE solutions.
These findings suggest that cheetahs achieve high-speed running while reducing leg loading by using a type-EG-like phase relationship between spinal movement and stance phases, thereby effectively exploiting spinal flexion.

% 本研究ではさらに，チーターの体幹剛性が曲げられているとき(flexed)と伸ばされているとき(extended)で非対称であることに着目し，その非対称性が走行に与える影響を調べた．
% 結果として，$\kappa$を大きくすることで高い速度を保ちながら床反力を低下させることは可能である一方で，安定性は損なわれるため，最適な$\kappa$は存在しないことが示された．
% チーターの背骨において体幹剛性は非対称性で，反らす方向に硬い(即ち$\kappa>1$)と考えられる．
% チーターはこのような非対称な特性のもとでも，type EGの解に相当する運動を用いることで，走行中に脚にかかる負担を低減させつつ高い進行速度を実現している可能性が示唆される．
% ただし，$\kappa$が大きすぎると解の安定性や存在領域が小さくなっていくため，チーターの背骨の剛性非対称性は，安定性と他のパフォーマンスが両立するように，何らかの評価関数のもとで最適な大きすぎない値を取っている可能性が示唆される．

In this study, we also focused on the asymmetry of cheetah spinal stiffness between flexed and extended postures and investigated how this asymmetry affects running.
As a result, we showed that although increasing $\kappa$ can reduce the GRF while maintaining high speed, it also degrades stability; therefore, no single optimal $\kappa$ exists.
In cheetahs, spinal stiffness is considered asymmetric, with greater stiffness in the extension direction (i.e., $\kappa>1$).
These results suggest that, even under such asymmetric properties, cheetahs may achieve high horizontal velocity while reducing leg loading during running by using a phase relationship between spinal movement and stance phases corresponding to type EG solutions.
However, when $\kappa$ becomes too large, solution stability and the solution-existence region become smaller.
Therefore, cheetah spinal-stiffness asymmetry may be controlled to a specific value under some objective function that balances stability and other performance measures.

% 本研究では，体幹剛性の非対称性をモデル化するために，体幹が曲げられているとき(flexed)と伸ばされているとき(extended)で剛性が異なり，伸ばされているときに剛性が大きいと仮定している
% このモデルは，チーターなどの4足動物は重力に抗って立ち，体幹部は内蔵を支える必要があることから，脊柱の構造が棘突起などの構造により背側（反らす方向）には動かしにくく，腹側（曲げる方向）には動かしやすくなっていることを反映している．
% さらに，後肢離地時に背筋で加速することにも起因して，脊柱の伸筋である軸上筋は，屈筋である軸下筋よりも筋肉の断面積が大きく大きな力を発揮しやすいという性質もモデル化している．
% これまでモデル研究やロボットへの応用において，このような非対称な剛性が用いられ，運動パフォーマンスを改善することが示されている\cite{Khoramshahi2013-hp, Zappetti2020-fv}．
% ただし，このように姿勢に依存する剛性の変化は，背骨の複雑な構造や筋腱の特性を十分に表したものではない．
% % ヒトに対する研究などで，背骨のflexionとextensionにおける剛性の違いが報告されているが\cite{Ghezelbash2018-pj,Wang2020-vb}，これらは筋のflexionとextensionにおける筋力の違いに起因するものであり，姿勢によって剛性が一意に定まるという本研究のモデルとは異なる．
% 今後の研究では，解剖学的な知見も導入して，体幹剛性の非対称性をより詳細にモデル化する必要がある．

In this study, to model asymmetric spinal stiffness, we assumed that stiffness differs between flexed and extended postures and is greater in the extended posture.
This model reflects the fact that, in quadrupeds such as cheetahs, animals must stand against gravity and support their internal organs; therefore, spinal structures such as spinous processes make dorsal motion (extension direction) harder and ventral motion (flexion direction) easier.
In addition, because acceleration at hindlimb liftoff is assisted by back muscles, the model also reflects the property that epaxial muscles, which are spinal extensors, have larger cross-sectional area and can generate larger forces more easily than hypaxial muscles, which are flexors.
Asymmetric stiffness of this kind has been used in previous model studies and robotic applications, where it has been shown to improve locomotor performance~\cite{Khoramshahi2013-hp, Zappetti2020-fv}.
However, posture-dependent stiffness changes of this type do not fully represent the complex structure of the spine or musculotendon properties.
% Although studies in humans and other subjects have reported differences in spinal stiffness between flexion and extension~\cite{Ghezelbash2018-pj,Wang2020-vb}, these differences are attributed to differences in muscle force between flexion and extension and therefore differ from our model, in which stiffness is uniquely determined by posture.
Future studies should incorporate anatomical findings to model spinal-stiffness asymmetry in greater detail.

\subsection{Limitations and future work}
% 本研究では脊柱の運動と前後肢の接地の位相関係でtype EGとGEの差異を明らかにした．
% ただし実際のチーターの走行においては、体幹の曲げ伸ばしだけでなく，脚の振り（外側に伸展するか、内側に屈曲するか）の違いも重要な役割を持つ．
% また，接地時も，脚は体幹に対して最も前方にあるのではなく，接地する前から足先は後方に移動し始め，接地直後の制動力を抑えている．
% 本研究で提案したモデルでは脚の力学が考慮されていないため上記の効果については議論できない．今後の研究において，脚の動力学と体幹の運動の相互作用について調べていきたい．

In this study, we clarified the differences between type EG and type GE based on the phase relationship between spinal motion and forelimb--hindlimb stance.
However, in actual cheetah running, not only spinal flexion--extension but also differences in leg swing (i.e., whether the limbs extend outward or flex inward) play important roles.
In addition, at touchdown, the leg is not at its most anterior position relative to the trunk; rather, the foot has already begun moving backward before touchdown, which suppresses braking force immediately after contact and helps maintain high-speed locomotion.
Because the model proposed in this study does not include leg dynamics, these effects cannot be addressed.
In future work, we will investigate the interaction between leg dynamics and spinal motion.

% 本研究では非対称な体幹剛性を姿勢依存で変化させたが，実際の動物では筋の活動(flexionとextensionにおける筋力の違い)に起因する剛性の変化があると考えられる．
% 加えて，実際の動物では剛性変化のしきい値にも非対称性があると予想される．
% 具体的には，少し曲げた状態が平衡状態となる可能性がある．
% 今後の研究では，解剖学的な知見も導入して，体幹剛性の非対称性をより詳細にモデル化する必要がある．

In this study, we modeled asymmetric spinal stiffness as a posture-dependent change.
However, in real animals, stiffness changes are likely to arise from muscle activity (i.e., differences in muscle force between flexion and extension).
In addition, the threshold for stiffness change is expected to be asymmetric in real animals.
Specifically, a slightly flexed posture may be the equilibrium state.
In future work, it will be necessary to model spinal-stiffness asymmetry in greater detail by incorporating anatomical findings.

% また今回用いたモデルは保存系であり，制御系が考慮できていない．
% 今後は神経系モデルや機械学習の導入を行い，身体制御の方法についても考察していきたい．

Moreover, the model used in this study is a conservative system, and control systems are not considered.
In the future, we would like to introduce neural models to investigate the body control methods.

\subsection*{Acknowledgments}
This study was supported in part by JSPS KAKENHI Grant Number JP24K17232 and by JST FOREST Program Grant Number JPMJFR2021.

\subsection*{Conflict of interest}
The authors declare no conflict of interest.

\subsection*{Data availability}
The data that support the findings of this study are available from the corresponding author upon reasonable request.

% bibtex
\bibliographystyle{IEEEtran}
\bibliography{2026_BB}

\end{document}